\pdfoutput=1
\documentclass[11pt]{article}
\usepackage[utf8]{inputenc}
\usepackage[top=70pt,bottom=70pt,left=60pt,right=60pt]{geometry}
\usepackage{amsmath}
\usepackage{hyperref}
\usepackage{amssymb}
\usepackage{cite}
\usepackage{ mathrsfs }
\usepackage[space]{grffile}
\usepackage{graphicx}
\newcommand{\overbar}[1]{\mkern 1.5mu\overline{\mkern-1.5mu#1\mkern-1.5mu}\mkern 1.5mu}

\begin{document}
\thispagestyle{empty}

{\hbox to\hsize{
\vbox{\noindent June 2019}}}

\noindent
\vskip2.0cm
\begin{center}

{\Large\bf No-scale supergravity with new Fayet-Iliopoulos term}

\vglue.3in

Yermek Aldabergenov~${}^{a,b}$
\vglue.1in

${}^a$~Department of Physics, Faculty of Science, Chulalongkorn University,\\
Thanon Phayathai, Pathumwan, Bangkok 10330, Thailand\\
${}^b$~Institute of Experimental and Theoretical Physics, Al-Farabi Kazakh National University, \\
71 Al-Farabi Avenue, Almaty 050040, Kazakhstan \\
\vglue.1in
yermek.a@chula.ac.th
\end{center}

\vglue.3in

\begin{center}
{\Large\bf Abstract}
\vglue.2in
\end{center}

We find a new class of $N=1$ no-scale supergravity models with F- and D-term supersymmetry breaking, using a new Fayet-Iliopoulos term. The minimal setup contains one $U(1)$ vector multiplet and one neutral chiral multiplet parametrizing $SL(2,\mathbb{R})/U(1)$ manifold, with constant superpotential and linear gauge kinetic function. In our construction the FI term is field-dependent, and one can obtain flat vanishing potential (Minkowski vacuum) with broken SUSY, and global $SL(2,\mathbb{R})$ invariance (self-duality) of the bosonic equations of motion. The spectrum of the model includes a massive spin-1/2 field as well as a vector, a scalar, and a pseudo-scalar -- all classically massless. We discuss several modifications/extensions of the model as well as the introduction of matter fields. We also find a two-field extension of already existing no-scale model.

\newpage

\section{Introduction}

Supergravity (SUGRA) is a viable phenomenological framework that on the one hand accommodates low-energy effective descriptions of more fundamental theories like superstrings, and on the other hand provides UV extensions of Supersymmetric Standard Model. A particular class of four-dimensional $N=1$ supergravity models called "no-scale supergravity" is known to address the hierarchy problem of the Standard Model by dynamically determining the energy scales of supersymmetry (SUSY) breaking and electroweak symmetry breaking, by radiative corrections \cite{Cremmer:1983bf,Ellis:1983sf,Ellis:1983ei,Ellis:1984bm} (for a review see e.g. Refs. \cite{Lahanas:1986uc,Nanopoulos:1994as}). No-scale SUGRA has two defining properties: (i) Minkowski (or de Sitter~\cite{Ellis:2018xdr}) vacuum with classically flat directions along which SUSY is spontanously broken, and (ii) vanishing ${\rm Str}{\cal M}^2$ \cite{Ellis:1983ei,Ellis:1984bm,Ferrara:1994kg}. The latter requirement has to be fulfilled once we include all possible fields of a realistic model in order to avoid quadratically divergent quantum corrections to the scalar potential.

Most of the discussions of no-scale SUGRA involve pure F-term SUSY breaking, where the minimal model contains a single chiral multiplet. However, in Ref. \cite{DallAgata:2013jtw} a new class of no-scale models was proposed where a chiral multiplet is coupled to a $U(1)$ vector multiplet, and SUSY is broken in Minkowski vacuum by auxiliary fields of both multiplets (F- and D-term SUSY breaking). The purpose of this work is to explore new no-scale scenarios with mixed F- and D-term breaking that emerged after the discovery of new Fayet-Iliopoulos terms \cite{Cribiori:2017laj,Kuzenko:2018jlz} (see also Ref. \cite{Kuzenko:2017zla} and Refs. therein for preceding developments, and Refs. \cite{Farakos:2018sgq,Aldabergenov:2018nzd,Aldabergenov:2017hvp,Antoniadis:2018cpq,Antoniadis:2018oeh,Cribiori:2018dlc,Abe:2018plc,Abe:2018rnu} for applications of new FI terms). Along the way we will consider a two-field extension of the model of Ref. \cite{DallAgata:2013jtw}.

This paper is organized as follows. In Sec. 2 we recall the form of the scalar potential of general $N=1$ SUGRA, and the simplest no-scale model with pure F-term SUSY breaking. In the first half of Sec. 3 we review the basic properties of the no-scale model of Ref. \cite{DallAgata:2013jtw} with mixed F- and D-term breaking, while in the second half we introduce the two-field extension of that model. In Sec. 4 we introduce a new class of no-scale models (with F- and D-term breaking) by using the new FI term. Sec. 5 is devoted to modifications and extensions of the simplest scenario, while in Sec. 6 we discuss general matter couplings in our models. Sec. 7 is left for conclusion and discussion.

\section{No-scale supergravity with F-term SUSY breaking}

The classical scalar potential of a general (standard) four-dimensional $N=1$ supergravity is a sum of F- and D-term potentials \cite{Wess:1992cp,Freedman:2012zz}:~\footnote{Throughout the paper we work in Planck units, $M_P=1$.}
\begin{gather}
    V_F=e^K\left[K^{i\bar{j}}(W_i+K_iW)(\overbar{W}_{\bar{j}}+K_{\bar{j}}\overbar{W})-3|W|^2\right]~,\label{VF}\\
    V_D=\frac{g^2}{2}f_R^{AB}\mathscr{D}_A\mathscr{D}_B~,\label{VD}
\end{gather}
where $K=K(\Phi_i,\overbar{\Phi}_i)$ is a K\"ahler potential depending upon general chiral (complex) scalar fields $\Phi_i$, $W=W(\Phi_i)$ is a holomorphic superpotential, $f=f(\Phi_i)$ is a holomorphic gauge kinetic function with $f_R\equiv {\rm Re}f$, $g$ is the gauge coupling, and $\mathscr{D}$ is a Killing potential, or moment map, of a given gauge group. As for the indices, we use the notation $K^{i\bar{j}}\equiv K_{i\bar{j}}^{-1}$ where $K_{i\bar{j}}\equiv\frac{\partial^2K}{\partial\Phi_i\partial\overbar{\Phi}_j}$, $W_i\equiv\frac{\partial W}{\partial\Phi_i}$, and $f^{AB}\equiv f_{AB}^{-1}$ where $A,B$ are gauge group indices. Killing potentials can be expressed as
\begin{equation}
    \mathscr{D}_A=i\left(K_i+\frac{W_i}{W}\right)X^i_A~,\label{Killingpot}
\end{equation}
where $X^i_A$ are Killing vectors of gauge symmetries. The last term in Eq. \eqref{Killingpot} is present only when R-symmetry is gauged.

The potential \eqref{VF}\eqref{VD}, as well as the full Lagrangian of $N=1$ SUGRA, is invariant w.r.t. the K\"ahler-Weyl transformations
\begin{equation}
    K\rightarrow K+\Sigma+\overbar{\Sigma}~,~~~W\rightarrow We^{-\Sigma}~,\label{KWtransform}
\end{equation}
where $\Sigma$ is an arbitrary chiral (super)field.

The essential part of the scalar potential that allows for Minkowski vacua with spontaneously broken SUSY is the negative contribution $-3|W|^2$ in Eq. \eqref{VF}.

The simplest no-scale model is described by the K\"ahler potential~\footnote{K\"ahler potentials of the form $K=-\alpha\log(T+\overbar{T})$ describe $SL(2,\mathbb{R})/U(1)\cong SU(1,1)/U(1)$ manifolds.} \cite{Cremmer:1983bf,Ellis:1983sf,Ellis:1983ei}
\begin{equation}
    K=-3\log(T+\overbar{T})~,\label{K3Log}
\end{equation}
depending on the single chiral field $T$, and the constant superpotential $W_0$. In this case the scalar potential \eqref{VF} vanishes due to the field identity
\begin{equation}
    K^{T\overbar{T}}K_TK_{\overbar{T}}=3~,
\end{equation}
and the whole theory has global $SL(2,\mathbb{R})$ invariance \cite{Ellis:1983ei}, where the complex scalar $\tau=i\overbar{T}$ transforms as
\begin{equation}
    \tau\rightarrow\frac{p\tau+q}{r\tau+s}~,~~~p,q,r,s\in\mathbb{R}~,~~~ps-qr=1~.\label{sl2r}
\end{equation}

Because the potential vanishes, for any value of $T$ there is a Minkowski vacuum where SUSY is broken by a VEV of the auxiliary field $F_T$ (the index here represents the multiplet to which the $F$-field belongs),
\begin{equation}
    \langle F_T\rangle=\langle -e^{K/2}K^{T\overbar{T}}(\overbar{W}_{\overbar{T}}+K_{\overbar{T}}\overbar{W})\rangle=\overbar{W}_0\langle T+\overbar{T}\rangle^{-1/2},
\end{equation}
with the gravitino mass given by
\begin{equation}
    \langle m_{3/2}\rangle=\langle e^{K/2}|W|\rangle=|W_0|\langle T+\overbar{T}\rangle^{-3/2}~.
\end{equation}
The goldstino (the superpartner of $T$) is absorbed by the gravitino, while the spectrum also includes two classically massless real scalars ${\rm Re}T$ and ${\rm Im}T$.

The K\"ahler potential \eqref{K3Log} can be accompanied by a cubic superpotential,
\begin{equation}
    W=\mu T^3~,~~~\mu\in\mathbb{C}~, \label{WT3}
\end{equation}
without spoiling the flatness of the potential \cite{Ellis:2018xdr}. By a K\"ahler-Weyl transformation \eqref{KWtransform} Eqs.~\eqref{WT3} and \eqref{K3Log} can be rescaled to
\begin{equation}
    W\rightarrow W=\mu~,~~~K\rightarrow K=-3\log\left(\frac{1}{T}+\frac{1}{\overbar{T}}\right)~,
\end{equation}
which merely amounts to a field redefinition, $T\rightarrow 1/T$.

\section{Models with F- and D-term SUSY breaking}

In Ref. \cite{DallAgata:2013jtw} it was shown that a no-scale model can accommodate both F- and D-term SUSY breaking. A particular model considered in the aforementioned work includes a chiral multiplet $\{T,\chi\}$ and a vector multiplet $\{A_m,\lambda\}$, where $T$ is a complex scalar, $\lambda$ and $\chi$ are spin-1/2 fields, and $A_m$ is a vector. The K\"ahler potential, superpotential, and gauge kinetic function of the model are given by
\begin{equation}
\begin{aligned}
    K&=-2\log(T+\overbar{T})~,\\
    W&=W_0~,~~~f=f_0~,
    \end{aligned}
\end{equation}
where $W_0$ and $f_0$ are complex constants. The vector field $A_m$ gauges the axionic shift symmetry
\begin{equation}
    T\rightarrow T+i\beta~,~~~\beta\in\mathbb{R}~.
\end{equation}
The corresponding Killing vector and Killing potential are
\begin{equation}
    X^T=i~,~~~\mathscr{D}=\frac{2}{T+\overbar{T}}~.
\end{equation}

This leads to the scalar potential $V=V_F+V_D$ with
\begin{equation}
    V_F=-\frac{|W_0|^2}{(T+\overbar{T})^2}~,~~~V_D=\frac{2g^2}{f^R_0(T+\overbar{T})^2}~,
\end{equation}
where $g$ is the gauge coupling. The potential vanishes for
\begin{equation}
    |W_0|^2=\frac{2g^2}{f_0^R}~,
\end{equation}
in which case both SUSY and the $U(1)$ gauge symmetry are spontaneously broken. The gravitino becomes massive by absorbing the goldstino -- a linear combination of $\chi$ and $\lambda$, and $A_m$ becomes massive by absorbing the pseudo-scalar ${\rm Im}T$. Another combination of $\chi$ and $\lambda$ appears as a massive fermion, while ${\rm Re}T$ remains massless.

The VEVs of the auxiliary scalars $F_T$ and $D$, and the gravitino mass are given by
\begin{gather}
    \langle F_T\rangle=\overbar{W}_0~,~~~\langle D\rangle=-\frac{2g}{\langle T+\overbar{T}\rangle}~,\\
    \langle m_{3/2}\rangle^2=\frac{|W_0|^2}{\langle T+\overbar{T}\rangle^2}=\frac{2g^2}{f_0^R\langle T+\overbar{T}\rangle^2}~.
\end{gather}

We would like to show now that the model can be non-trivially extended by adding a neutral chiral field $S$ as
\begin{gather}
    K=-2\log(T+\overbar{T})-\log(S+\overbar{S})~,\\
    W=W_0+\mu S~,
\end{gather}
where $\mu$ is a complex parameter, and gauge kinetic function is constant as before $f=f_0$.

The shift symmetry of $T$ is gauged, and the scalar potential is given by
\begin{equation}
    V_F=-\frac{\mu\overbar{W}_0+\overbar{\mu}W_0}{(T+\overbar{T})^2}~,~~~V_D=\frac{2g^2}{f_0^R(T+\overbar{T})^2}~.
\end{equation}
If
\begin{equation}
    \mu\overbar{W}_0+\overbar{\mu}W_0<\frac{2g^2}{f^R_0}~,
\end{equation}
we have a quasi de Sitter vacuum (runaway in the ${\rm Re}T$ direction) with two flat directions: ${\rm Re}S$ and ${\rm Im}S$. If
on the other hand
\begin{equation}
    \mu\overbar{W}_0+\overbar{\mu}W_0=\frac{2g^2}{f^R_0}~,
\end{equation}
the potential vanishes and we have a Minkowski vacuum.

In both cases SUSY is broken by non-vanishing auxiliary fields,
\begin{gather}
    \langle F_T\rangle=\frac{\overbar{W}_0+\overbar{\mu}\langle\overbar{S}\rangle}{\langle S+\overbar{S}\rangle^{1/2}}~,~~~\langle F_S\rangle=(\overbar{W}_0-\overbar{\mu}\langle S\rangle)\frac{\langle S+\overbar{S}\rangle^{1/2}}{\langle T+\overbar{T}\rangle}~,\\
    \langle D\rangle=-\frac{2g}{\langle T+\overbar{T}\rangle}~,
\end{gather}
while the gravitino mass is
\begin{equation}
    \langle m_{3/2}\rangle^2=\frac{|W_0+\mu \langle S\rangle|^2}{\langle T+\overbar{T}\rangle^2\langle S+\overbar{S}\rangle}~.
\end{equation}
As in the previous case the $U(1)$ symmetry is broken at the minimum and ${\rm Im}T$ is absorbed by the massive vector boson.

\section{No-scale model with new FI term}

New Fayet-Iliopoulos terms that do not require gauging of R-symmetry were proposed in Refs. \cite{Cribiori:2017laj,Kuzenko:2018jlz}. We choose one of them that in curved superspace can be expressed as (we use the conventions of Ref. \cite{Wess:1992cp})
\begin{equation}
    {\cal L}_{\rm FI}=g\xi\int d^2\Theta 2{\cal E}\overbar{\cal P}\left(\frac{{\cal W}^2\overbar{\cal W}^2}{{\cal P}{\cal W}^2\overbar{\cal P}\overbar{\cal W}^2}{\cal DW}\right)+{\rm h.c.}~,\label{superfieldFI}
\end{equation}
where $\xi$ is the real FI constant, $g$ is the gauge coupling, and $\cal P$ is the chiral projector ${\cal P}\equiv{\cal D}^2-8\overbar{{\cal R}}$, $\overbar{\cal P}\equiv\overbar{\cal D}^2-8{\cal R}$; the (chiral) superfield strength of the vector superfield $\cal V$ is defined as ${\cal W}_\alpha\equiv -\frac{1}{4}\overbar{\cal P}{\cal D}_\alpha {\cal V}$ with ${\cal W}^2\equiv {\cal W}^\alpha {\cal W}_\alpha$. In the absence of chiral matter the FI term \eqref{superfieldFI} leads to the positive constant scalar potential $V_D=g^2\xi^2/2$, i.e. de Sitter vacuum. This potential comes from eliminating the auxiliary field $D$, but not from a Killing potential $\mathscr{D}$.

This new FI term allows us to introduce another class of no-scale models in $N=1$ SUGRA that can be described by the superspace Lagrangian
\begin{equation}
    {\cal L}=\int d^2\Theta 2{\cal E}\left\lbrace\overbar{\cal P}\left[e^{-K/3}\left(\frac{3}{8}+g\xi\frac{{\cal W}^2\overbar{\cal W}^2}{{\cal P}{\cal W}^2\overbar{\cal P}\overbar{\cal W}^2}{\cal DW}\right)\right]+W+\frac{f}{4}{\cal W}^2\right\rbrace+{\rm h.c.}~,\label{superfieldL}
\end{equation}
where $K$, $W$, and $f$ are K\"ahler potential, superpotential, and gauge kinetic function respectively.~\footnote{In superspace actions a generic chiral field $\Phi$ is promoted to a chiral superfield, and $K(\Phi,\overbar{\Phi})$, $W(\Phi)$, and $f(\Phi)$ should be understood as functions of superfields.}

The minimal setup with a chiral field $S$ is given by
\begin{gather}
    \begin{gathered}
        K=-\log(S+\overbar{S})~,\\
        W=W_0~,~~~f=aS~,\label{MAK}
    \end{gathered}
\end{gather}
where $W_0$ is a complex constant, and $a$ is a positive real constant. In this case the bosonic components of the Lagrangian \eqref{superfieldL} (after eliminating the auxiliary fields and rescaling to Einstein frame) read
\begin{equation}
    e^{-1}{\cal L}=\frac{1}{2}R-\frac{\partial_m S\partial^m\overbar{S}}{(S+\overbar{S})^2}-\frac{a}{8}(S+\overbar{S})F_{mn}F^{mn}-\frac{ia}{8}(S-\overbar{S})F_{mn}\Tilde{F}^{mn}-V~,\label{compL}
\end{equation}
where $F_{mn}$ is the field strength of the abelian vector $A_m$ belonging to the vector superfield $\cal V$, and $\tilde{F}^{mn}\equiv\frac{1}{2}\epsilon^{mnpq}F_{pq}$ with Levi-Civita tensor $\epsilon$. The scalar potential $V=V_F+V_D$ is given by
\begin{equation}
    V_F=-\frac{2|W_0|^2}{S+\overbar{S}}~,~~~V_D=\frac{g^2\xi^2}{a(S+\overbar{S})}~.\label{VFInoscale}
\end{equation}

In terms of the canonical scalar $\phi$ and the pseudo-scalar $C$, which we will refer to as dilaton and axion, the chiral field $S$ can be parametrized as $S=e^{-\sqrt{2}\phi}+iC$.

The model admits flat vanishing potential (Minkowski vacuum) with spontaneously broken supersymmetry when
\begin{equation}
    |W_0|^2=\frac{g^2\xi^2}{2a}~.\label{WFIcond}
\end{equation}
The auxiliary fields develop non-vanishing VEVs,
\begin{equation}
    \langle F_S\rangle=\overbar{W}_0\langle S+\overbar{S}\rangle^{1/2}~,~~~\langle D\rangle=-\frac{2g\xi}{a\langle S+\overbar{S}\rangle}~,
\end{equation}
and the gravitino mass is given by
\begin{equation}
    \langle m_{3/2}\rangle^2=\frac{|W_0|^2}{\langle S+\overbar{S}\rangle}=\frac{g^2\xi^2}{2a\langle S+\overbar{S}\rangle}~.
\end{equation}

Due to the fermionic structure of the new FI term ($D$ appears in the denominator of a gravitino-gaugino-vector coupling) $\xi$ must be non-zero \cite{Cribiori:2017laj}, and therefore SUSY is necessarily broken. Also, because 
\begin{equation}
    {\cal P}{\cal W}^2=2F_{mn}F^{mn}+2iF_{mn}\tilde{F}^{mn}-4D^2+{\rm fermions}+{\cal O}(\Theta)~
\end{equation}
must be nowhere vanishing \cite{Cribiori:2017laj,Kuzenko:2018jlz}, as can be seen from Eqs. \eqref{superfieldFI} and \eqref{superfieldL}, we have the condition on the field strength
\begin{equation}
    F_{mn}F^{mn}+iF_{mn}\tilde{F}^{mn}-2D^2\neq 0~.\label{Fbound}
\end{equation}

In the absence of the scalar potential, the action \eqref{compL} has global $SL(2,\mathbb{R})$ invariance, similar to that of the single-field model \eqref{K3Log}, where the complex scalar $\tau=i\overbar{S}$ transforms as in Eq. \eqref{sl2r} (the constant $a$ can be eliminated by the rescaling of the vector field, $A_m\rightarrow A_m/\sqrt{a}$). However, in the case of the action \eqref{compL} the field strength $F_{mn}$ also non-trivially transforms under this group \cite{Montonen:1977sn}, and the $SL(2,\mathbb{R})$ invariance becomes an extension of $SO(2)$ electromagnetic self-duality. It is worth mentioning that the $SL(2,\mathbb{R})$ is symmetry of the equations of motion rather than the action \eqref{compL} itself. Taking all this into account we can enforce the vanishing of the scalar potential \eqref{VFInoscale} (via the relation \eqref{WFIcond}) by requiring the $SL(2,\mathbb{R})$ invariance of the equations of motion.

The tree-level spectrum of the model includes a massive spin-1/2 field, massless vector, and two massless real scalars.

\section{Modifications and extensions of the model}

The model of Eq. \eqref{MAK} can be modified as
\begin{gather}
    \begin{gathered}
        K=-3\log(S+\overbar{S})~,\\
        W=W_0+\mu S~,\label{MBK}
    \end{gathered}
\end{gather}
where we added a linear term with the complex parameter $\mu$ to the superpotential and changed the numerical factor in the K\"ahler potential to "$-3$", while gauge kinetic function $f=aS$ is unchanged. The corresponding scalar potential reads
\begin{gather}
    V_F=-\frac{2|\mu|^2}{S+\overbar{S}}-\frac{\mu\overbar{W}_0+\overbar{\mu}W_0}{(S+\overbar{S})^2}~,\label{MBVF}\\
    V_D=\frac{g^2\xi^2}{a(S+\overbar{S})}~.\label{MBVD}
\end{gather}
Requiring Minkowski vacuum then leads to the following conditions on the parameters
\begin{gather}
    \mu\overbar{W}_0+\overbar{\mu}W_0=0~,\label{MBmuWcond}\\
    |\mu|^2=\frac{3g^2\xi^2}{2a}~.\label{MBmuFIcond}
\end{gather}

As a simple example consider the case where $\mu$ is real. Then, Eq. \eqref{MBmuWcond} forces $W_0$ to be pure imaginary, and as a consequence the superpotential is restricted to the form (using Eq. \eqref{MBmuFIcond})
\begin{equation}
    W=ib\pm\sqrt{\frac{3}{2a}}g\xi S~,
\end{equation}
where $b$ is an arbitrary real constant.

Interestingly, the model \eqref{MAK} can also be modified to accommodate multiple "no-scale" chiral fields. We find that the two-field setup
\begin{gather}
    \begin{gathered}
        K=-\log(S+\overbar{S})-2\log(T+\overbar{T})~,\\
        W=W_0+\mu T~,~~~f=aS~,\label{MCK}
    \end{gathered}
\end{gather}
leads to vanishing scalar potential provided that
\begin{gather}
    \mu\overbar{W}_0+\overbar{\mu}W_0=0~,\label{MCmuWcond}\\
    |\mu|^2=\frac{2g^2\xi^2}{a}~.\label{MCmuFIcond}
\end{gather}

Similarly, the three-field case is given by
\begin{gather}
    \begin{gathered}
        K=-\sum_{i=1}^{3}\log(\Phi_i+\overbar{\Phi}_i)~,\\
        W=W_0+\mu T+\rho Y~,~~~f=aS~,\label{MDK}
    \end{gathered}
\end{gather}
where $\Phi_i={S,T,Y}$ are the three chiral fields, and $W_0$, $\mu$, $\rho$ are complex parameters.

The conditions for vanishing potential are
\begin{gather}
    \mu\overbar{W}_0+\overbar{\mu}W_0=\rho\overbar{W}_0+\overbar{\rho}W_0=0~,\label{MDmurhoWcond}\\
    \mu\overbar{\rho}=\overbar{\mu}\rho=\frac{g^2\xi^2}{2a}~.\label{MDmuFIcond}
\end{gather}

\section{Matter couplings}

For phenomenological applications any no-scale model should be coupled to general matter fields in a way that doesn't destabilize Minkowski or de Sitter vacua and preserve flat direction(s).

Considering the minimal model (Eq. \eqref{MAK}) of our framework, the most general matter couplings are of the form~\footnote{Our matter couplings are similar to those described in Ref. \cite{DallAgata:2013jtw} if we set $Q=0$.}
\begin{gather}
    K=-\log(S+\overbar{S}-Q)+{\cal K}(S,\overbar{S},\Phi_i,\overbar{\Phi}_i)~,\\
    W=W_0+\Omega(\Phi_i)~,\\
    f=aS+h(\Phi_i)~,
\end{gather}
where $Q=Q(S,\overbar{S},\Phi_i,\overbar{\Phi}_i)$ and $\cal K$ are real functions of general matter fields $\Phi_i$ and the "no-scale" field $S$, while $\Omega$ and $h$ are holomorphic functions of $\Phi_i$.

In order to preserve no-scale structure of the model, these functions must vanish at the minimum,
\begin{equation}
    \langle Q\rangle=\langle{\cal K}\rangle=\langle\Omega\rangle=\langle h\rangle=0~.\label{vevpotentials}
\end{equation}
Then, taking into account Eq. \eqref{WFIcond}, the full scalar potential reads
\begin{multline}
    \langle V\rangle=\langle S+\overbar{S}\rangle^{-1}\left\langle K^{Sj^*}K_SW_0(\overbar{\Omega}_{j^*}+K_{j^*}\overbar{W}_0)+K^{i\overbar{S}}(\Omega_i+K_iW_0)K_{\overbar{S}}\overbar{W}_0+\right.\\\left.+K^{ij^*}(\Omega_i+K_iW_0)(\overbar{\Omega}_{j^*}+K_{j^*}\overbar{W}_0)\right\rangle~.\label{vevmatter}
\end{multline}
Thus, to obtain Minkowski vacuum the quantity in the large angle brackets must vanish. For example, if we assume that $Q=h=0$, and ${\cal K}=\Phi_i\overbar{\Phi}_i+{\rm higher~order~corrections}$, the vanishing of Eq. \eqref{vevmatter} is guaranteed if $\Omega$ contains terms at least quadratic in $\Phi_i$ (the presence of linear terms in $\Omega$ would shift the minimum so that $\langle{\cal K}\rangle\neq 0$ in general).

Since our models have a massless vector field in the spectrum, for phenomenological purposes it is desirable to give it a mass by Higgs or St\"uckelberg mechanism, so that it would decouple at lower energies. To do so, we can gauge a $U(1)$ isometry of, say, $\cal K$ by coupling one of the chiral fields, $Z\in\Phi_i$, to the vector field $A_m$, and require (in addition to the vanishing of Eqs. \eqref{vevpotentials} and \eqref{vevmatter}) that $\langle X^Z\rangle\neq 0$, where $X^Z$ is the corresponding Killing vector.

\section{Discussion and conclusion}\label{concl}

Our two main results of this work are the following: (i) we found a two-field extension of the no-scale supergravity model with mixed F- and D-term supersymmetry breaking proposed in Ref. \cite{DallAgata:2013jtw}, and (ii) we introduced a new class of no-scale supergravity models by using one of the new FI terms that do not require gauging of R-symmetry \cite{Cribiori:2017laj,Kuzenko:2018jlz}. In particular, we showed that the negative F-term potential in our setup can be cancelled by the positive (dilaton-dependent) FI term, leading to Minkowski vacuum with flat directions in several different scenarios including models with multiple "no-scale" chiral fields. In these cases the conditions for Minkowski vacuum are given by specific relations between the FI constant $\xi$, the coefficient of the gauge-kinetic function $a$, and the parameters of a given superpotential. In our models the vanishing of the potential can be enforced by requiring the global $SL(2,\mathbb{R})$ invariance of the equations of motion. But since we considered only the bosonic actions of these models, it needs to be checked whether or not the full theories can be made $SL(2,\mathbb{R})$-invariant by assigning appropriate transformation rules for dilatino and gaugino. Nonetheless it is clear that the fermionic terms of the model \eqref{MBK} are not $SL(2,\mathbb{R})$-invariant because the superpotential depends on the pseudo-scalar ${\rm Im}S$ (recall that under $SL(2,\mathbb{R})$ the ${\rm Im}S$ shifts by a constant real number) which will couple to (superpotential-dependent) fermionic terms, and thus the $SL(2,\mathbb{R})$ transformations will change the equations of motion.

The study of our models can be continued in several directions. First, the FI term \eqref{superfieldFI} that we used is just one representative of a whole family of new FI terms as was shown in Ref. \cite{Kuzenko:2018jlz}. An alternative choice for example would be \cite{Kuzenko:2018jlz}
\begin{equation}
    {\cal L}_{\rm FI}'=g\xi'\int d^2\Theta 2{\cal E}\overbar{\cal P}\frac{{\cal W}^2\overbar{\cal W}^2}{({\cal D}{\cal W})^3}+{\rm h.c.}~,\label{superfieldFI2}
\end{equation}
One of the differences with our previous choice is that this FI term does not impose restrictions on the field strength (see Eq. \eqref{Fbound}). Next, the models we proposed may be of use in the context of cosmology, namely for the description of inflation and/or dark energy. And finally, since these are purely phenomenological models, as we mentioned in Introduction, they should be derivable from a more fundamental framework like superstrings/M-theory in order to make contact with quantum gravity.

\section*{Acknowledgements}

Y.A. was supported by the CUniverse research promotion project of Chulalongkorn University under the grant reference CUAASC, and the Ministry of Education and Science of the Republic of Kazakhstan under the grant reference AP05133630.

\bibliography{Bibliography.bib}{}
\bibliographystyle{utphys.bst}

\end{document}